\documentclass[openacc]{rstransa}

\usepackage{graphicx}
\usepackage{amsmath}
\usepackage{natbib}
\def\Planck{\it Planck\rm}
\def\LCDM{$\Lambda$CDM}
\def\apj{ApJ }
\def\apjl{ApJL }
\def\apjs{ApJS }
\def\nat{Nature }

\def\aap{A\&A }
\def\aapr{Astr. Astrophys. Reviews }
\def\prd{Phys. Rev. D  }
\def\prl{Phys. Rev. Lett.  }
\def\mnras{MNRAS }
\def\jcap{J. Cosmology Astropart. Phys. }
\def\Hunit{${\rm km} \  {\rm s}^{-1} {\rm Mpc}^{-1}$ }
\def\Hunitns{${\rm km} \ {\rm s}^{-1} {\rm Mpc}^{-1}$}
\newcommand{\Plik}{\textit{Plik}}
\newcommand{\CamSpec}{\textit{CamSpec}}

\def\pmb#1{\setbox0=\hbox{#1}%
    \kern-.025em\copy0\kern-\wd0
    \kern.05em\copy0\kern-\wd0
    \kern-.025em\raise.0433em\box0}

\def\ltsima{$\; \buildrel < \over \sim \;$}
\def\gtsima{$\; \buildrel > \over \sim \;$}
\def\simlt{\lower.5ex\hbox{\ltsima}}
\def\simgt{\lower.5ex\hbox{\gtsima}}

\titlehead{Challenges to $\Lambda$CDM}

\begin{document}

\title{Challenges to the $\Lambda$CDM Cosmology}

\author{
George Efstathiou}

\address{Kavli Institute for Cosmology, \\  Madingley Road, \\ Cambridge, CB3 OHA.}

\subject{Cosmology}


\corres{George Efstathiou\\
\email{gpe@ast.cam.ac.uk}}

\begin{abstract}
  Observations of the cosmic microwave background (CMB) radiation are described to remarkable accuracy
  by the six-parameter $\Lambda$CDM cosmology. However, the key ingredients of this model,
  namely dark matter, dark energy and cosmic inflation are not understood at a fundamental level. It is therefore important to investigate tensions between the CMB and other cosmological probes. I will review aspects of tensions with direct measurements of the Hubble constant $H_0$, measurements of weak gravitational lensing, and the recent hints of evolving dark energy reported by the Dark Energy Spectroscopic Instrument (DESI) collaboration.
\end{abstract}


\begin{fmtext}
  \section{Introduction}
  \label{sec:Introduction}

The field of cosmology has been fortunate in having a major satellite mission dedicated to
measuring the CMB in each of the last three decades.  NASA's Cosmic Background Explorer (COBE)
led to  the discovery of CMB anisotropies in 1992  \citep{Smoot:1992}.  NASA's Wilkinson Microwave Anisotropy Probe,  released its first results in 2003 \citep{Bennett:2003} and established the six parameter \LCDM\ cosmology in the form that we know it today. The ESA \Planck\ satellite was launched in 2009 and the final results from the \Planck\ collaboration were presented in 2018  \citep{Params:2020}. An overview  paper  summarizing the cosmological legacy of the \Planck\ mission \citep{Legacy:2020} concluded: '{\it The 6-parameter $\Lambda$CDM model continues to provide an excellent fit to the cosmic microwave background data at high and low redshift, describing the cosmological information in over a billion map pixels with just six parameters}'. 

In this contribution to the Royal Society Discussion Meeting, I will review whether the above quotation is still justified today. I will assume throughout that the Universe is homogeneous and isotropic on large scales,  since this is a key ingredient of the $\Lambda$CDM cosmology and

\end{fmtext}

\maketitle

\noindent
is supported by observations of the CMB.  It is, of course, important to test the assumptions of homogeneity and isotropy of our Universe using different types of data. Such tests are described by others at this meeting.

\section{The Exquisite fit of \LCDM\ to CMB anisotropies}
\label{sec:CMB}

\begin{figure}[!h]
\centering\includegraphics[width=5.0in]{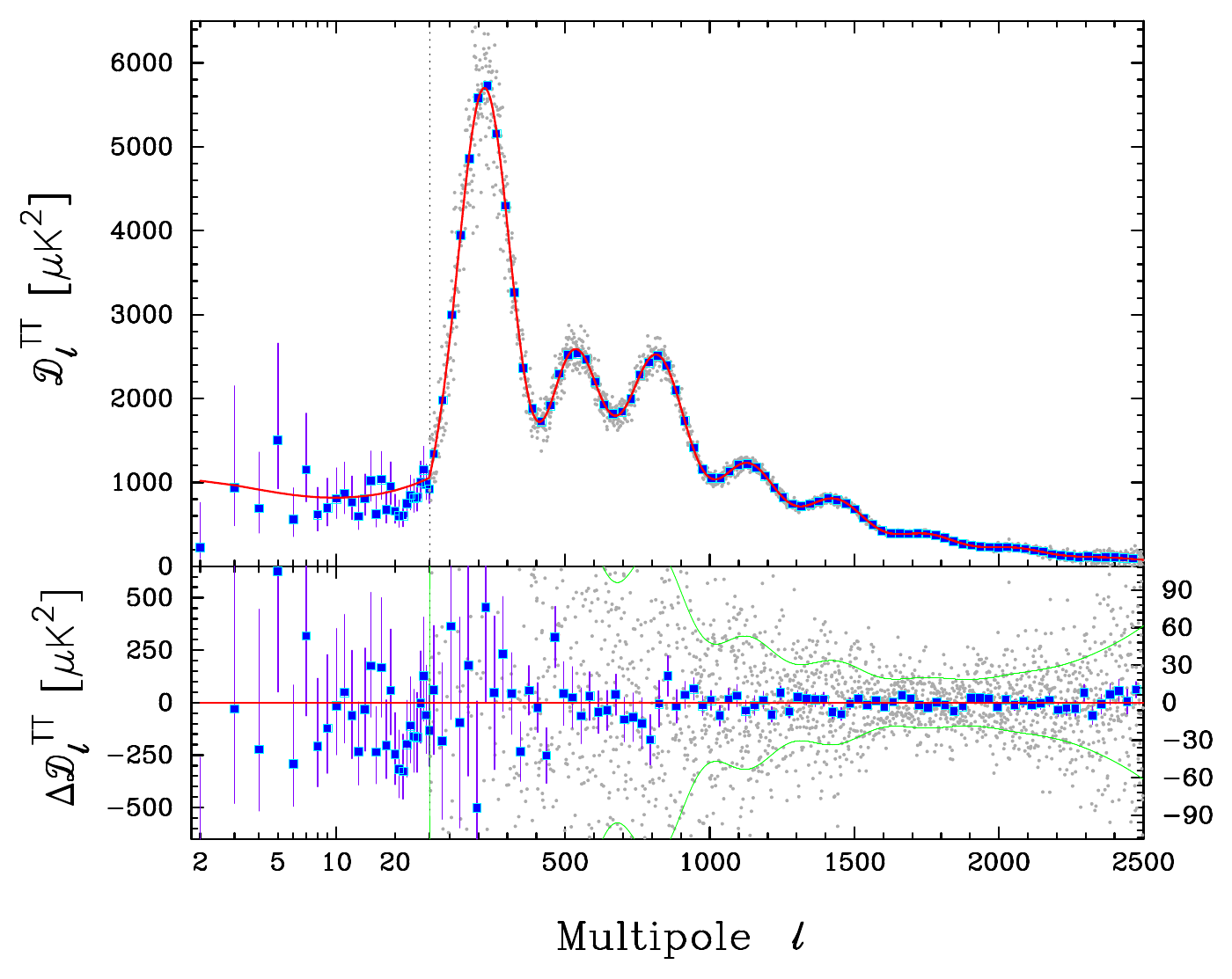}
\caption{The upper panel shows the \Planck\ CMB temperature power spectrum and the lower panel shows the residuals with respect to the power spectrum of the base six parameter \LCDM\ model fitted to the TTTEEE spectra (shown by the red line in the upper panel).  The multipole scale is logarithmic over the multipole range $2-29$ and linear
  at higher multipoles. The power spectrum computed over 86\% of the sky from the 
{\tt Commander} component separated maps \cite{Likelihood_2020} is shown over the multipole range $2-29$ together with asymmetrical 68\% error bars. The foreground corrected frequency average power spectrum computed from the NPIPE \citep{NPIPE} \Planck\ maps, averaged in multipole bins of  width $\Delta \ell= 30$,   are plotted as the blue points. The faint grey points show the power spectrum  multipole-by-multipole. The error bars show $1\sigma$ errors on the band powers computed from the diagonals of the high multipole covariance matrix. The green lines plotted in the lower panel show the $\pm 1\sigma$ error ranges for the grey points. } 
\label{fig:TT_spec}
\end{figure}

\begin{figure}[!h]
\centering\includegraphics[width=5.2in]{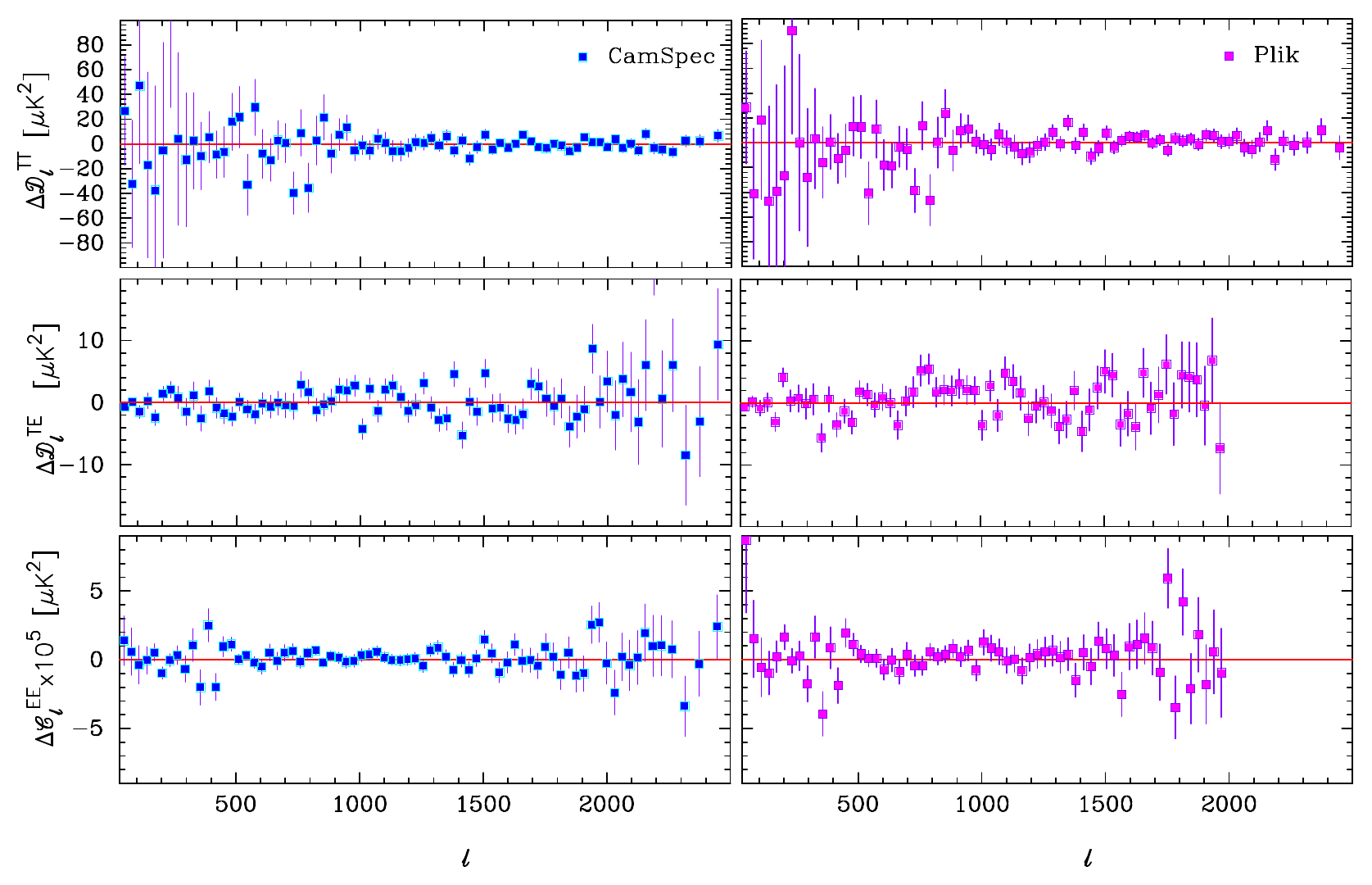}
\caption{
  Residuals of the TT, TE and EE spectra relative to the best fit \LCDM\ model
  plotted in Fig. \ref{fig:TT_spec}. Residuals for NPIPE \CamSpec\ spectra are shown in the left hand panels. Residuals  for the \Plik\ spectra,  as used in the baseline  2018 \Planck\ TTTEEE likelihood, relative to the same cosmology are shown in the right hand panels. (Adapted from \cite{Efstathiou:2023}).}
\label{fig:CL_residuals}
\end{figure}

Following the 2018 \Planck\ data release,  in collaboration with Steven Gratton and Erik Rosenberg I began a programme 
to extract more information from the \Planck\ power spectra  by modifying the
\CamSpec\ pipeline \citep{Params:13} to use larger sky areas and dust cleaned spectra
\citep{Efstathiou:2021, Rosenberg:2022}. The foreground corrected, frequency averaged,  temperature power spectrum from the most recent iteration \citep{Efstathiou:2023} based on the NPIPE \Planck\ maps  \citep{NPIPE} is shown in Fig. \ref{fig:TT_spec}. The high multipole power spectrum shown in this plot is effectively a full mission average of the $143\times 143$, $143\times217$ and $217\times 217$ power spectra computed over
80\% of the sky. The cosmological parameters of the base six parameter \LCDM\ cosmology computed from this analysis are almost identical to those reported in \citep{Params:2020}. However, the residuals with respect to the best fit model are substantially smaller. This is true for each of the TT, TE and EE spectra as illustrated in Fig. \ref{fig:CL_residuals}.

This is a significant result. In our attempts to extract more
information from \Planck, the temperature and polarization spectra
lock on even more accurately to the base \LCDM\ cosmology.  There is
no evidence for any new physics beyond \LCDM. Indications of  anomalies,
such as the excess smoothing of the TT acoustic peaks (quantified by the
phenomenological parameter $A_L$), variations in cosmological parameters as a function
of multipole \citep{Addison:2016} and features in the
\Planck\ spectra \citep[e.g.][]{Obied:2017}, all  decrease in statistical significance, 
consistent with the behaviour expected of statistical fluctuations. This conclusion is strengthened further by the extremely good agreement between the \Planck\ \LCDM\ best fit model and the TE and EE spectra extending to high multipoles measured by the Atacama Cosmology Telescope (ACT) and by the South Pole Telescope (SPT) \citep{Choi:2020, Dutcher:2021}.

If it is argued that  new physics beyond \LCDM\ is required to explain, 
for example,  distance scale measurements of the Hubble
constant (the 'Hubble tension'), galaxy weak lensing measurements, (the '$S_8$
tension'), evolving dark energy, or bulk flow anomalies (as described elsewhere in this
volume)  then that new physics must reproduce the CMB anisotropies of the
base \LCDM\ cosmology to extraordinarily high precision. Additional data beyond CMB measurements therefore
become critical in assessing such challenges to \LCDM.

\maketitle
\section{The Hubble tension}
\label{sec:Hubble}

Fitting the base \LCDM\ model to the TTTEEE spectra of Figs. \ref{fig:TT_spec} and \ref{fig:CL_residuals} combined with the {\tt Commander} TT and {\tt SIMALL} EE likelihoods at $\ell < 30$ \citep{Likelihood_2020} (I will use the terminology \Planck\ TTTEEE to  refer to this combination of the high and low multipole likelihoods), we find a Hubble constant of
\begin{equation}
H_0 = 67.43 \pm 0.49 \ {\rm km} \  {\rm s}^{-1} {\rm Mpc}^{-1}. \label{equ:H01}
\end{equation} 
In contrast, the latest value of the Hubble constant measured by the SH0ES collaboration based on Cepheid variables and Type Ia
supernovae (SN) is 
\begin{equation}
H_0 = 73.01 \pm 0.99 \ {\rm km}\  {\rm s}^{-1} {\rm Mpc}^{-1},  \label{equ:H02}
\end{equation}
\citep{Riess:2022}. These two numbers differ by more than $5\sigma$, a discrepancy that has become known as the 'Hubble tension'
\citep[for recent reviews see][]{Shah:2021, Freedman:2021, Tully:2023,  Breuval:2024}.

It is extremely unlikely that the \Planck\ value of \ref{equ:H01} is
wrong. There is a huge degree of redundancy in the \Planck\ data and
so there are many different ways in which the data can be
partitioned. For example, the TE spectra alone (which are free of
extragalactic foregrounds) give an $H_0$ consistent with
Eq. \ref{equ:H01} and with comparable accuracy. Furthermore, the high
resolution ground based experiments give independent estimates of the
Hubble parameter for the \LCDM\ cosmology that are consistent with the
\Planck\ value and differ from the SH0ES value by many standard
deviations ($H_0 = 67.6 \pm 1.1$ \Hunit for ACT TTTEEE combined with
WMAP, \citep{Aiola:2020}; $H_0 = 68.3 \pm 1.5$ \Hunit for SPT=3G
TTTEEE, \citep{Balkenhol:2023}).  It is therefore reasonable to
conclude that either the \LCDM\ model is missing new physics or the
SH0ES estimate is biased in some way. Freedman, in this volume,
presents new JWST observations of Cepheids, tip of the red giant
branch, and carbon-rich asymptotic giant branch stars to infer $H_0 =
69.8 \pm 1.9$ \Hunitns, slightly lower than the SH0ES value and
consistent with the CMB value. On the other hand, JWST Cepheid
photometry by the SH0ES team is in very good agreement with their
earlier HST results, effectively eliminating systematics associated
with crowded field photometry as the source of the tension
\citep{Riess:2023, Riess:2024}. Evidently more work needs to be done
to achieve a consensus between Freedman and collaborators and the
SH0ES team. For the rest of this section I will take the SH0ES result
at face value and discuss whether it is possible to modify \LCDM\ to
explain their value of $H_0$.

\begin{table}[!h]
\caption{Values of the Hubble constant with $1\sigma$ errors  for extensions to  \LCDM.}
\label{tab:variants}
\begin{center}
\begin{tabular}{lll}
\hline
Model & \Planck\ TTTEEE & \Planck\ TTTEEE+BAO \\
\hline
\LCDM\  &  $67.44 \pm 0.58$ & $67.69 \pm 0.42$ \\
\LCDM + $m_\nu$ & $66.8 \pm 1.2$ & $67.8 \pm 0.6$ \\
\LCDM + $N_\nu$ & $66.4 \pm 1.6$ & $67.4 \pm 1.2$ \\
\LCDM + $m_\nu+N_\nu$ & $66.1^{+1.9}_{\ -1.6}$  & $67.5 \pm 1.2$ \\
\LCDM + $m_{\rm str}+N_\nu$ & $67.1 \pm 0.7$ & $67.89^{+0.45}_{-0.69}$ \\
\LCDM + $n_{\rm run}$ &  $67.25 \pm 0.6$  & $67.66 \pm 0.45$ \\
\LCDM + $\Omega_k$ & $56 \pm 4$ & $67.9 \pm 0.7$ \\
\LCDM + $w_0+w_a$ &  --   & $64.9 \pm 2.1$ \\ \hline

\end{tabular}
\end{center}
\vspace*{-4pt}
\end{table}
  
To begin,  Table \ref{tab:variants} shows the posteriors for $H_0$ for variants of \LCDM\ from the grid of models discussed in 
\cite{Params:2020}. $m_\nu$ is the  mass of a single massive neutrino eigenstate  
(fixed to $0.06 \ {\rm eV}$ in the base model, as expected for a normal hierarchy), $N_\nu$ is the
number of neutrino/neutrino-like relativistic species (fixed to 3.046 in the base model), $m_{\rm str}$ adds a massive sterile 
neutrino, $n_{\rm run}$ adds a running of the scalar spectral index,  $\Omega_k$ is the spatial curvature and the parameters
$w_0$ and $w_a$ model  dynamical dark energy (see Sec. \ref{sec:desi}). In the latter two variants, the CMB anisotropies 
suffer a large geometrical degeneracy \citep{Efstathiou:1999} and cannot determine $H_0$ accurately. Thus the third column adds baryon acoustic oscillation (BAO) measurements as described in \cite{Params:2020} to break the geometrical  degeneracy. {\it There is not even a hint of movement towards the SH0ES value of $H_0$ in any of these variants}.

As discussed in Sec.~\ref{sec:CMB}, the \Planck\ power spectra  are extremely well
fit by the base \LCDM\ cosmology. The $H_0$ posteriors of these variants peak at values close to
that of base \LCDM, but additional model complexity can introduce degeneracies which increase the error in $H_0$. Almost all proposed cosmological 'solutions' to the $H_0$ tension \citep[see][for a review]{diValentino:2021}
are of this type, i.e. favouring an $H_0$ that is close to the value of base \LCDM\ but increasing the error because of internal parameter degeneracies. Furthermore, because the CMB is so well fit by base \LCDM, the interpretation of theoretical solutions
to the $H_0$ tension becomes sensitive to the use, and sometimes misuse,  of supplementary astrophysical data \citep[see e.g.][]{Efstathiou:2021b}.

It is well known that by combining BAO measurements, the magnitude-redshift relation of type Ia SN,   and the CMB value  of the sound
horizon $r_d$, it is possible to construct  an inverse distance ladder for $H_0$ \citep[see e.g.][]{Heavens:2014, Aubourg:2015, Abbott:2018}. In fact, it is not even necessary to assume  \LCDM\ since BAO and Type Ia SN strongly constrain the background expansion history to be close to that of \LCDM\ irrespective of dynamics \citep{Heavens:2014, Verde:2017,Lemos:2019}. Modifications to \LCDM\ at low redshift,
for example adding interactions between dark matter and dark energy, dynamical dark energy, or decaying dark matter cannot significantly affect the inverse distance ladder. 
  For example, \cite{Lemos:2019} assume the \Planck\ value $r_d = 147.27 \pm 0.31 \ {\rm Mpc}$ and use BAO measurements together with  the Pantheon SN sample \citep{Scolnic:2018} to infer,  in a model independent way, 
\begin{equation}
H_0 = 68.42 \pm 0.88 \ {\rm km} \  {\rm s}^{-1} {\rm Mpc}^{-1}, \ {\rm inverse \ distance \ ladder}, \ \ \Planck \  r_d , \label{equ:IL1}
\end{equation}
which is in tension with SH0ES at about the $3.5 \sigma$ level. This suggests that the Hubble tension requires a mechanism  to 
lower the sound horizon \citep[see e.g.][]{Knox:2020}. However, the \LCDM\ model is remarkably consistent.  For example,  \cite{Brieden:2023} bypass the CMB estimate of $r_d$, using
BAO measurements and constraints from big bang nucleosynthesis (BBN) to infer $H_0 = 67.42^{+0.86}_{-0.94}$ \Hunitns, while \cite{Philcox:2022}
use full-shape galaxy power spectrum (sensitive to the scale of matter-radiation equality) with \Planck\ CMB lensing and
Type Ia SN measurements to infer  $H_0 = 64.8^{+2.2}_{-2.5}$ \Hunitns, again without assuming the \Planck\ value for $r_d$.

Early dark energy (EDE) is an attempt to preserve the physics of \LCDM\ at both high and low redshift, by introducing a 'confusiton' -- a
scalar field
$\phi$  that is dynamically important only at around the time of recombination \citep[see the reviews by][and references therein]{Kamionkowski:2023,Poulin:2023}.  Here I will review some results from \cite{Efstathiou:2023} in which we
 considered a scalar field evolving in an axion-like potential:
\begin{equation}\label{eq:potential}
    V(\theta) = m^2 f^2[1-\cos (\theta)]^3,
\end{equation} where $m$ represents the axion mass, $f$ the axion decay constant, and $\theta \equiv \phi/f$ is a re-normalized field variable defined such that $-\pi \leq \theta \leq \pi$.  This potential had been considered by many authors \citep[e.g.][]{Poulin:2019, Smith:2020} and provides a flexible model with which to illustrate the observational consequences of EDE. This model adds three parameters to base \LCDM\  which (following \citet{Smith:2020})  we choose to be 
the critical redshift $z_c$ at which the 
scalar field starts to roll, the fractional contribution of EDE to the total energy density at that redshift 
$f_{\rm EDE}(z_c)$ and the initial field value $\theta_i$.

\begin{figure}[!h]
\hspace{-0.3in}\includegraphics[width=5.5in]{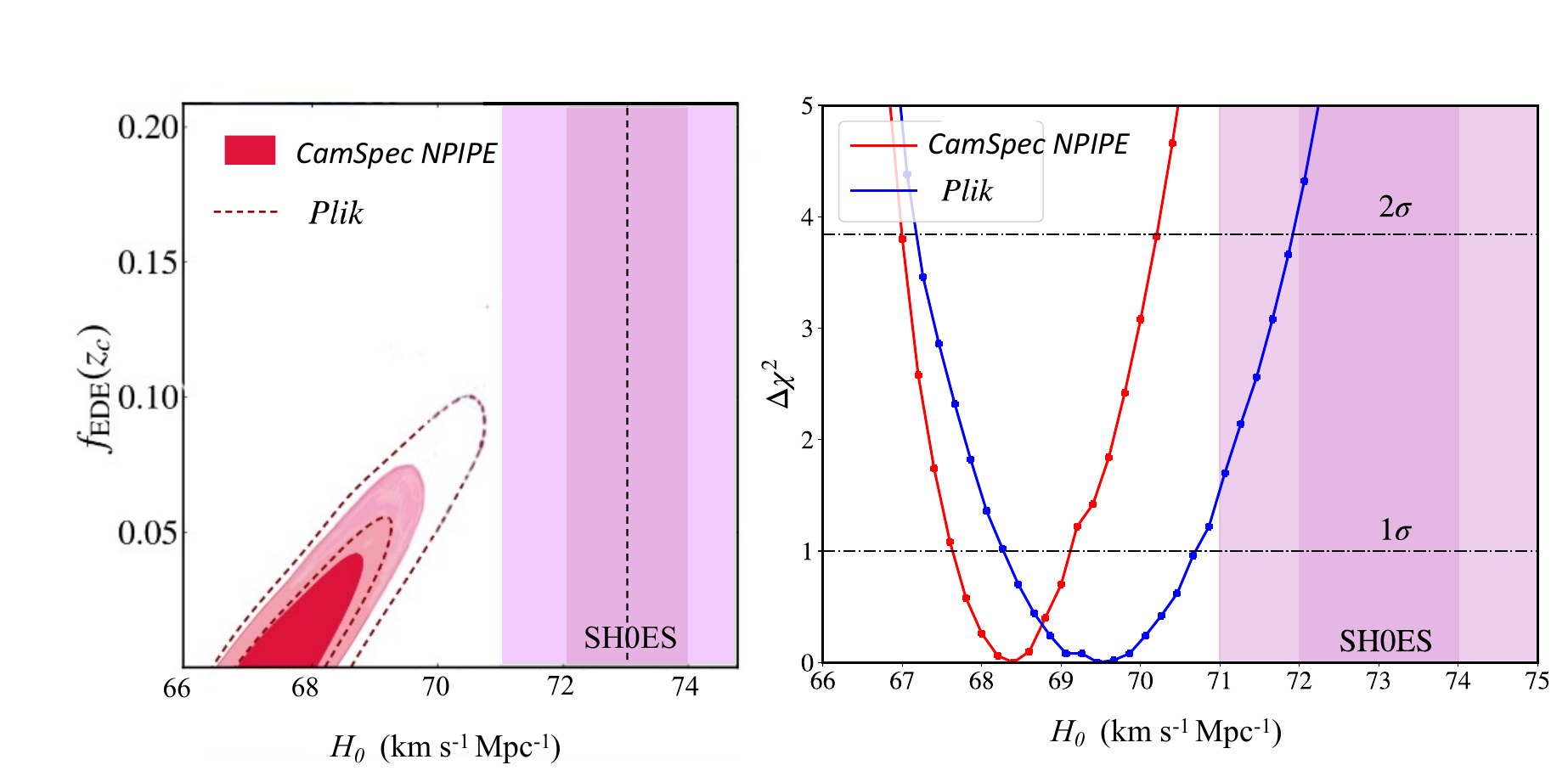}
\caption{Left hand plot shows Bayesian constraints on $H_0$ and the EDE parameter $f_{\rm EDE}(z_c)$ using two \Planck\ likelihoods in combination with BAO and SN data as described in the text. The right hand plot shows profile likelihoods of $H_0$. The purple shaded regions show the $1$ and $2\sigma$ ranges of the SH0ES measurement of $H_0$. (Adapted from \cite{Efstathiou:2023}).}
\label{fig:EDE}
\end{figure}

The left hand panel of Fig. \ref{fig:EDE} shows  Bayesian constraints on the
parameters $f_{EDE}(z_c)$ and $H_0$ fitted to the \Planck\ data
combined with BAO and SN data from the updated Pantheon+ SN catalogue
\citep{Scolnic:2022}. For details of the data used see
\cite{Efstathiou:2023}. The red contours show constraints obtained
using the \CamSpec\ NPIPE likelihood at multipoles $\ge 30$ (with
power spectrum residuals shown in Fig. \ref{fig:CL_residuals}) and the
dotted contours show the constraints obtained using the 2018
\Plik\ likelihood in place of \CamSpec. The right hand panel shows
profile likelihoods of $H_0$ which are independent of priors
\citep[see e.g][]{Herold:2022}. Both \Planck\ likelihoods disfavour
EDE and are in tension with the SH0ES value of $H_0$.  The key conclusion to be drawn from Fig. \ref{fig:EDE}
is that an improvement in the \Planck\ likelihood leads to even greater tension
with the SH0ES value of $H_0$ and favours parameters close to those of 
base \LCDM\ (irrespective of statistical methodology and choice of priors). EDE as a solution of the Hubble
tension is quite strongly disfavoured by the data. Similar conclusions have been reached
by \cite{McDonough:2023} and \cite{Qu:2024b} using different data combinations.

In summary, observational data probing both high and low redshifts set such strong constraints that it is
difficult to construct a plausible theory\footnote{It is possible to construct contrived models to evade these problems. For example it has been suggested that a sudden transition in the value of the  gravitational constant
within the last 100 million years might have led to a dimming of nearby Type Ia SN \cite[e.g.][]{Marra:2021}.} to match the SH0ES value of $H_0$ \citep[see also][]{Vagnozzi:2023}.
Any such theory must involve parameter degeneracies that combine fortuitously to mimic the 6-parameter \LCDM\ cosmology to high accuracy. This is why I regard the Hubble tension as such a frustrating challenge to \LCDM.

\section{The $S_8$ tension}
\label{sec:S8}

Surveys of weak galaxy lensing provide measures of the parameter combination\footnote{Where $\Omega_{\rm m}$ is the present day matter density in units of the critical density, $\sigma_8$ is the root mean square linear  amplitude of the matter fluctuation spectrum in spheres of radius $8 h^{-1} {\rm Mpc}$ extrapolated to the present day, and $h$ is the value of the Hubble constant $H_0$ in units of $100 \ {\rm km} \ {\rm s}^{-1}{\rm Mpc}^{-1}$.} $S_8 = \sigma_8 (\Omega_{\rm m}/0.3)^{0.5}$,  consistently finding values that are lower than that  expected according to  the \textit{Planck} best fit \LCDM\ cosmology.  This discrepancy has become known as the '$S_8$ tension'. 

\begin{figure}[!h]
\centering\includegraphics[width=4.0in]{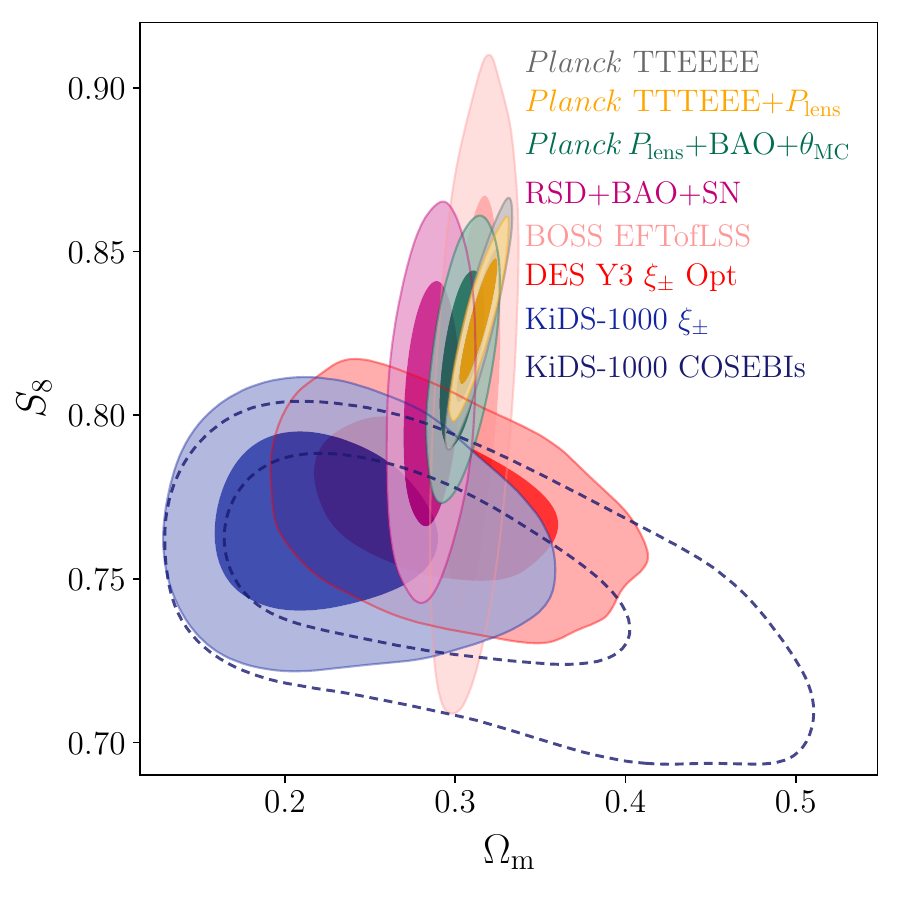}
\caption{68\% and 95\% constraints in the $S_8-\Omega_{\rm m}$ plane
  for various data assuming the 6-parameter \LCDM\ cosmology. The blue
  and navy (dashed) show the constraints from the KiDS $\xi_{\pm}$ and
  COSEBI statistics as analyzed by \cite{Asgari:2021}, while the red
  shows that from the DES Y3 \LCDM\ optimised $\xi_{\pm}$ analysis. The
  yellow and grey contours show constraints from \Planck\ TTTEEE with
  and without the addition of the \Planck\ CMB lensing likelihood
  (Plens). The peach contours labelled EFTofLSS show  constraints
  from the BOSS power spectrum and bispectrum effective field theory
  analysis of \citet{DAmico:2024}. The magenta contours show
  constraints from redshift space distortions (RSD) combined with BAO
  and SN measurements as described in \cite{Amon:2022}. The green
  contours show the constraint from the \Planck\ lensing likelihood
  combined with BAO together with conservative priors on the acoustic
  peak location parameter $\theta_{\rm MC}$ (an approximation to the parameter
$\theta_*$ defined in Sec \ref{sec:desi})  and other cosmological
  parameters.}
\label{fig:S8}
\end{figure}

In addition to weak lensing, there are other ways of measuring the
amplitude of the fluctuation spectrum. The perception has arisen that
$S_8$ tension reflects a difference between early time measures of the
fluctuation spectrum such as the CMB and late time probes \citep[see
  e.g.][]{Abdalla:2022}. Figure \ref{fig:S8} from \cite{Amon:2022}
challenges this perception. The plot shows constraints in the
$S_8-\Omega_m$ plane from the Kilo-Degree Survey (KiDS) and Dark Energy
Survey (DES) cosmic shear surveys. These contours sit low compared to
the constraints from \Planck\ (grey contours). The green contours show the constraints
from \Planck\ lensing combined with BAO and the CMB acoustic peak
location parameter $\theta_{\rm MC}$. CMB lensing is caused by matter
along the line of sight with a median redshift of $z \sim 2$, yet
$S_8$ is consistent with the value inferred from the primary
anisotropies. There is no evidence of a departure from the
\LCDM\ fluctuation growth rate between $z \sim 1000$ and $z \sim
2$. Furthermore, there is no evidence for 'gravitational slip';
photons are responding to the same gravitational potential as the
matter. The purple contours show constraints from redshift space
distortions (RSD) using the same galaxy and quasar survey RSD measurements  as
those used in \cite{Params:2020}. These surveys cover the redshift
range $0.1 - 1.5$, overlapping in  redshift with the  weak
lensing surveys. 
Although the errors are quite large, RSD are consistent with the primary 
CMB results with  no evidence for a
slowing of the linear growth rate with redshift.

 Several groups have developed `full-shape' analyses based on
 effective field theory (EFT) descriptions of non-linear perturbations
 \citep[e.g.][]{dAmico:2020, Ivanov:2020, Chen:2022, Philcox:2022b,
   DAmico:2024}.  The EFT analyses aim to constrain cosmological
 parameters independently of \textit{Planck}. However, the nuisance
 parameters required to model perturbation theory,
 galaxy biasing, and redshift space distortions, effectively
 down-weight information at wavenumbers $k \simgt 0.2 h^{-1} {\rm
   Mpc}$. As a consequence of the restricted wavenumber range, the
 primordial spectral index $n_{\rm s}$ is poorly constrained in
 comparison to \textit{Planck}. At the $\sim 1-2 \sigma$ level, EFT
 RSD depend on the choices of priors, particularly the parameter
 $n_s$. The peach coloured contours in Fig. \ref{fig:S8} show results
 from the \citep{DAmico:2024} EFT power spectrum and bispectrum
 analysis of the Baryon Oscillation Spectroscopic Survey (BOSS) galaxy
 sample. These authors apply a simulation based correction for prior
 volume effects in the EFT analysis that bias $\sigma_8$ low by about
 $1\sigma$ if left uncorrected. The errors in $S_8$ from this analysis
 are large but, significantly, there is no evidence of tension with
 the \Planck\ \LCDM\ cosmology. The role of priors in RSD analyses is
 discussed further by \cite{Brieden:2022, Maus:2023, Holm:2023}.

 Figure \ref{fig:S8} suggests an alternative interpretation of the
 $S_8$ tension \citep{Amon:2022}. The CMB and RSD measurements probe
 spatial scales that are in the linear regime. In contrast, weak
 lensing measurements are dominated by scales that are highly
 non-linear. If physical processes suppress the amplitude of the
 non-linear spectrum on small scales over and above the expectations of a universe composed of collisionless matter, then it may be possible to
 explain the $S_8$ tension while preserving the predictions of the
 base \LCDM\ cosmology on large scales where the density field is linear.  Baryonic feedback is
 an obvious mechanism that could produce such a suppression \citep[see e.g.][and
   references therein]{vanDaalen:2011, Vogelsberger:2014,
   Chisari:2019}. A more speculative proposal, which should not be discounted,  is that a suppression is caused by new 
physics in the dark sector, for example, a
 contribution from light axions
 \citep[see e.g.][and references therein]{Vogt:2023}. 
 
 \cite{Amon:2022}
 investigated the effects of power spectrum suppression by introducing
 the phenomenological model:
\begin{equation}
P_{\rm m}(k, z) =  P^{\rm L}_{\rm m}(k, z) + A_{\rm mod}[P^{\rm NL}_{\rm m} (k, z) - P^{\rm L}_{\rm m}(k, z)] \,. \label{equ:NL}
\end{equation} 
Here $P_{\rm m}(k, z)$ is the matter power spectrum at wavenumber $k$ and redshift $z$, the superscript  ${\rm L}$ denotes the 
linear theory power spectrum and  NL denotes the dark matter non-linear power spectrum  with no baryonic feedback 
(as computed, for example, by Euclid Emulator,  \cite{EuclidEmulator} or  {\sc HMCode2020},  \cite{Mead:2021}).  The parameter
$A_{\rm mod}$ is a constant that describes suppression of the spectrum on non-linear scales if $A_{\rm mod} < 1$. This model, and its relationship to models of baryonic feedback are described in greater detail in the talk by Amon at this meeting. As summarized by \cite{Preston:2023}, the \Planck\ \LCDM\ cosmology gives acceptable fits to the KiDS and DES Y3 cosmic shear data
for values of $A_{\rm mod}$ in the range $\sim 0.75 \ \mbox{--} \ 0.9$ depending on whether scale cuts  are  applied to the two-point statistics. The question of whether such values of $A_{\rm mod}$ can result from baryonic feedback is  controversial \citep[see e.g.][]{McCarthy:2023, Chen:2023, Arico:2023, Bigwood:2024} and so further work is required to assess whether the \cite{Amon:2022} proposal is viable. However, {\it it would only take one convincing measurement of a low value of $S_8$ on linear scales to falsify the proposal}.

\begin{figure}[!h]
\hspace{0.3in}\includegraphics[width=5.2in]{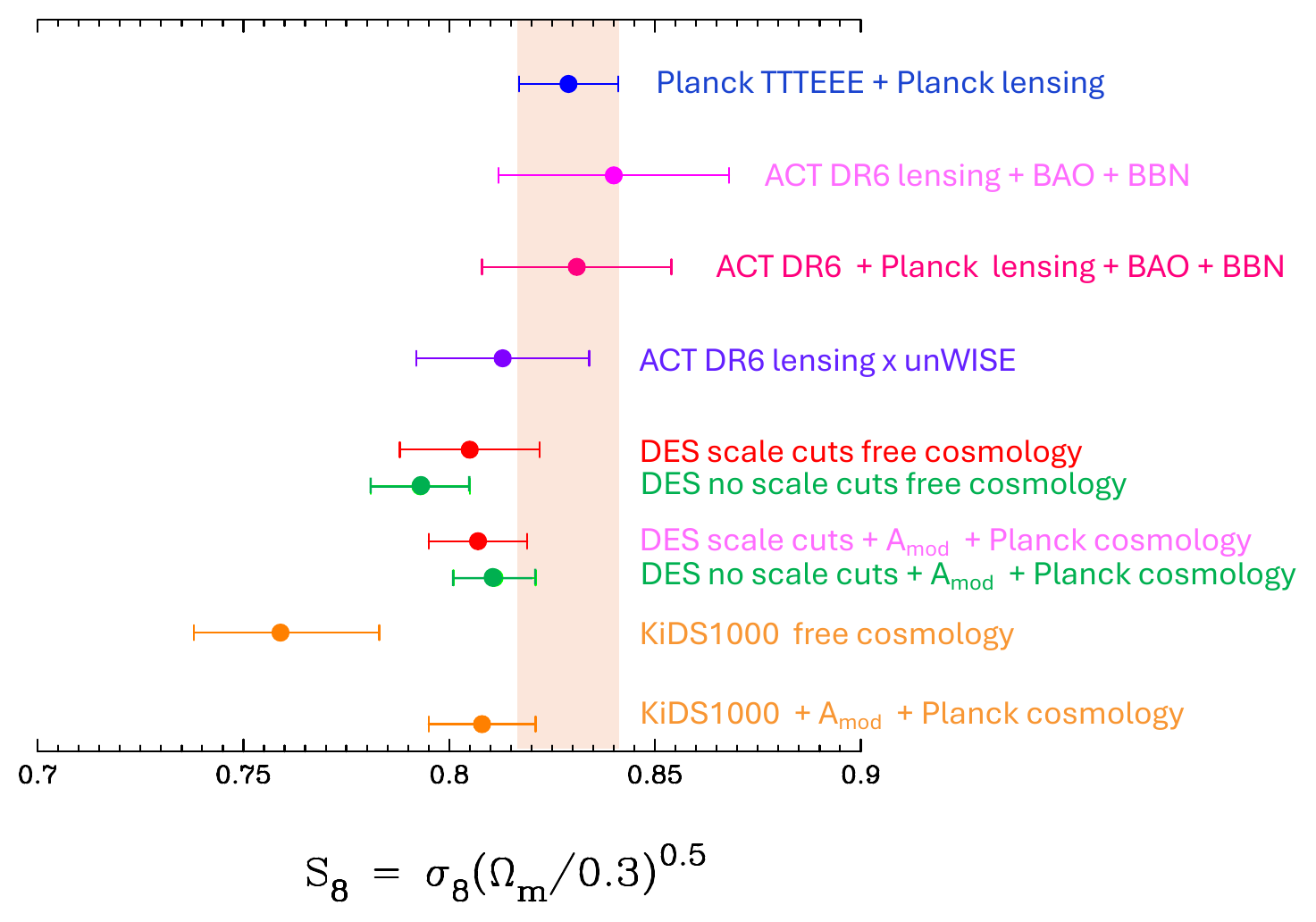}
\caption{Summary of measurements of $S_8$ including new results from ACT DR6 CMB lensing measurements \citep{Qu:2024b} and ACT DR6 lensing cross-correlate with unWISE galaxies
\citep{Farren:2024}. The remaining entries show results for the DES and KiDS   weak lensing surveys as  described in the text. }
\label{fig:S8summary}
\end{figure}

Figure \ref{fig:S8summary} includes two new measurements of $S_8$ on
linear scales. The points labelled ACT DR6 show new results on CMB
lensing from ACT Data Release  6 as described by \cite{Qu:2024}. These
measurements are consistent with the \Planck\ TTTEEE constraints on
$S_8$ and with \Planck\ lensing. The point labelled ACT DR6$\times$unWISE
shows the result of cross-correlating ACT DR6 lensing with galaxies
from the unWISE catalogue spanning the redshift range $z \sim 0.2-1.6$
\citep[see][for details]{Farren:2024}. Results from ACT DR6 are
discussed in greater detail by Madhavacheril at this meeting (including preliminary
results from cross-correlating ACT DR6 lensing with DESI luminous red galaxies (LRG)\footnote{Which can be compared to 
the cross-correlation analysis of DESI LRG with \Planck\ lensing which showed hints of a low value of $S_8$ 
\citep{White:2022}.}.

The remaining entries show results from \citep{Preston:2023} who
re-analyzed DES Y3 and KiDS-1000 $\xi_{\pm}$ cosmic shear
measurements including the parameter $A_{\rm mod}$. The red points in Fig. \ref{fig:S8summary} apply the
\LCDM\ optimized scale cuts to the DES Y3 $\xi_{\pm}$ measurements. As
discussed in \cite{Amon:2022} the DES Y3 analysis applied angular
scale cuts to reduce biases caused by baryonic feedback using the
EAGLE \citep{McAlpine:2016} and OWLS-AGN \citep{vanDaalen:2011}
hydrodynamic simulations as a reference
\citep[see][]{Krause:2021}. The green points labelled 'no scale cuts'
use all of the DES Y3 $\xi_{\pm}$ data points with angular separation $\ge
2.5^\prime$.  KiDS-1000 \citep{Asgari:2021} make more aggressive use
of small scales, retaining all scales with $\theta \ge 0.5^\prime$ in
$\xi_{+}$ and $\theta \ge 4^\prime$ in $\xi_{-}$. For each survey and choice of
scale-cuts we show results for $S_8$ allowing
cosmological parameters to vary with uninformative priors (labelled
'free cosmology'), with a \Planck\ \LCDM\ prior on cosmological
parameters (labelled '\Planck\ cosmology') and with and without
including the power spectrum suppression parameter $A_{\rm
  mod}$. Applying scale cuts to DES Y3 we see that allowing $A_{\rm
  mod}$ (with best fit value $A_{\rm mod} = 0.92 \pm 0.10$) to vary has
little effect. The weak lensing measurements on large angular scales
are consistent with the \Planck\ \LCDM\ cosmology.  However, if the
small scale data are included, consistency of DES Y3 data with the
\Planck\ \LCDM\ cosmology requires a suppression of the non-linear
power spectrum with $A_{\rm mod} = 0.86\pm 0.05$.  The KiDS-1000 data
probe even smaller scales and require $A_{\rm mod} = 0.75 \pm 0.07$ to
match the \Planck\ cosmology. If the suppression is interpreted in terms of  baryonic feedback, 
the latter two values of $A_{\rm mod}$  imply substantially stronger feedback than 
expected from recent hydrodynamical simulations \citep[e.g.][]{McCarthy:2017, McCarthy:2023}.

Clearly an accurate model of the non-linear matter power spectrum, including
the amplitude and scale dependence of the effects of baryonic feedback, is required 
to infer an unbiased value of $S_8$ from cosmic shear surveys\footnote{Differences  in the amplitude of the $S_8$ tension reported in the literature 
\citep{Asgari:2021, Arico:2023, Abbott:2023, Terasawa:2024} are caused, in part,  
by differences in  scale cuts and in the modelling of baryonic feedback.}. 
In the future
it will become possible to constrain baryonic feedback empirically using cross-correlations of
weak lensing and measurements of the kinetic and thermal Sunyaev-Zeldovich effects 
 \citep[e.g.][]{troester:2021, to2024, Bigwood:2024}. Such studies  will also provide valuable additional 
constraints on the modelling of feedback in numerical hydrodynamic simulations. RSD measurements
from DESI should provide a decisive test on whether the fluctuation growth rate at late times is
compatible with the \Planck\  \LCDM\ cosmology. 

\section{Evolving Dark Energy?}
\label{sec:desi}

Just before this meeting, BAO measurements from the first year of DESI
observations were submitted to the archive \citep{DESI_1:2024,
  DESI_2:2024}. Combining these BAO measurements with CMB observations
and various SN catalogues, the DESI team report evidence for a time
varying equation of state \citep{DESI_3:2024}, though they caution: '{\it it is important to
thoroughly examine unaccounted-for sources of systematic uncertainties
or inconsistencies between the different datasets that might be
contributing to these results}'. The DESI results are discussed by 
Palanque-Delabrouille at this meeting. Here I will make some general remarks
on whether the DESI results pose a challenge for \LCDM.

The DESI team parameterize the evolution of the dark energy equation-of-state (EoS) with redshift $z$
 as:
\begin{equation}
  w(z) = w_0 + w_a \left ( {z \over 1+z} \right ) ,  \label{equ:EoS1}
\end{equation}
introducing two additional parameters, $w_0$ and $w_a$ to the base \LCDM\  cosmology. This parameterization has the virtue 
of simplicity \citep{Linder:2003} but since the redshift dependence is constrained at both high and low redshift, one
must be cautious about interpreting constraints in $w_0-w_a$ space in terms of phantom crossing points ({\it i.e.} transitions of the EoS to phantom-like behaviour with $w(z) < -1$. This is because observational data constrain the EoS over a limited redshift range. The behaviour of $w(z)$ in the model of Eq.\ref{equ:EoS1} outside that limited range is a consequence of the parameterization rather than the data \citep[see e.g.][]{Wolf:2023, Cortes:2024, Shlivko:2024}. We will therefore treat Eq. \ref{equ:EoS1}  as a purely phenomenological parameterization.

\begin{figure}[!h]
\includegraphics[width=2.8in]{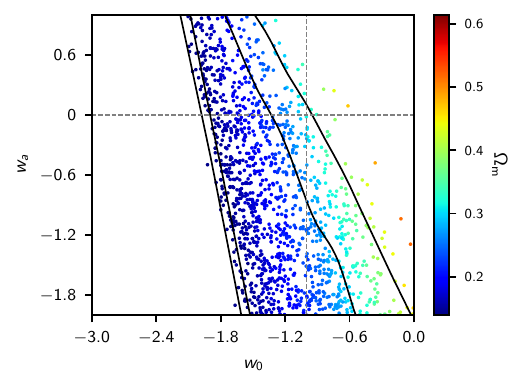}\includegraphics[width=2.8in]{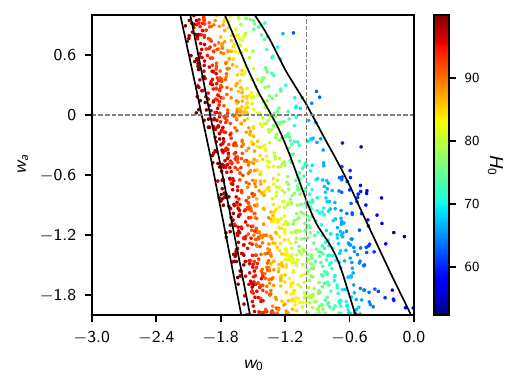}
\caption{Contraints on $w_0-w_a$ from the NPIPE \Planck\ TTTEEE likelihood colour coded by the value of $\Omega_m$ (left hand plot) and $H_0$ (right hand plot, with $H_0$ in units of \Hunitns). This plot shows that the CMB provides very weak constraints on $w_0-w_a$ because of a large geometrical degeneracy.}
\label{fig:CMB_w0wa}
\end{figure}

\begin{figure}[!h]
\hspace{-0.1in}\includegraphics[width=2.9in]{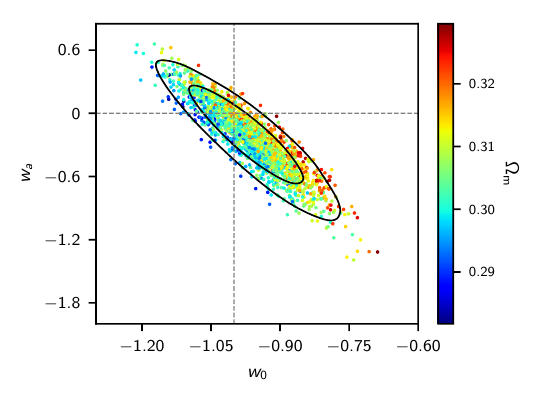}\hspace{-0.15in}\includegraphics[width=2.9in]{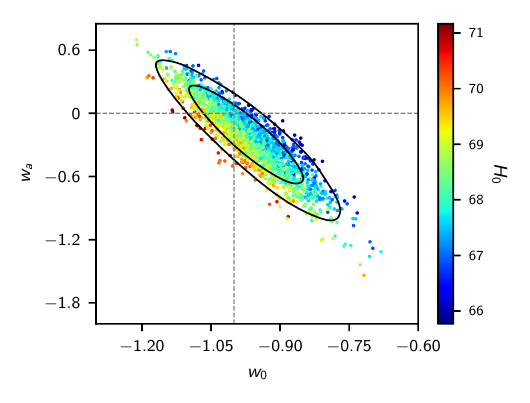}
\caption{Constraints on $w_0-w_a$ for \Planck\ TTTEEE combined with pre-DESI BAO measurements
(as used in \citep{Params:2020}) combined with the Pantheon SN catalogue. The MCMC samples
are colour coded by the value of $\Omega_m$ (left hand plot) and $H_0$ (right hand plot, with $H_0$ in units of \Hunitns). The cosmological constant corresponds to the intersection of the two dotted lines. }
\label{fig:CMB+BAO+SN_w0wa}
\end{figure}

Figure \ref{fig:CMB_w0wa} shows the constraints from \Planck\ TTTEEE
colour coded by $\Omega_m$ and $H_0$. The parameters $w_0, w_a$ are
highly degenerate, reflecting the geometrical degeneracy $\theta_* =
r_*/D_M(z_*)$, where $\theta_*$ is the acoustic peak location parameter
$r_*$ is the sound horizon at the time of recombination $z_*$ and 
$D_M(z_*)$ is the comoving angular diameter distance to $z_*$. The lines
in Fig. \ref{fig:CMB_w0wa} show the geometrical degeneracy $r_*/D_M(z_*)={\rm constant}$ for
fixed values of $\Omega_m$ and $H_0$. To maintain the acoustic peak structure, 
$\omega_b = \Omega_b h^2$, $\omega_c = \Omega_c h^2$  (and hence $\omega_m = \omega_b + \omega_c$) must be approximately constant, so the spread in 
Fig. \ref{fig:CMB_w0wa} is set approximately  by the error in $\omega_m$ for the base \LCDM\ cosmology.
This is why high values of $H_0$ in the right hand panel correspond to low values of $\Omega_m$ in the
left hand panel.  For $w_a=0$, the MCMC samples are skewed to values $w_0 < -1$. This is caused mainly by the 
slight TT power deficit at $\ell \simlt 30$ compared to the best fit base \LCDM\ model (see Fig. \ref{fig:TT_spec}) 
which tends to pull $w_0$ into the phantom domain via the integrated Sachs-Wolfe effect, though at low statistical
significance \citep{Escamilla:2024}. Figure \ref{fig:CMB_w0wa} shows why $H_0$ is unconstrained by 
 \Planck\ TTTEEE  if  $w_0, w_a$ are added as parameters to \LCDM\ (cf. Table \ref{tab:variants}).
Evidently CMB anisotropies alone  cannot constrain $w_0, w_a$. Any evidence for evolving dark energy must therefore come
from supplementary data.

Figure \ref{fig:CMB+BAO+SN_w0wa} adds BAO data, as described in \citep{Params:2020} (dominated statistically by BOSS DR12)
and the Pantheon SN sample. As noted in \citep{Params:2020}, there is no evidence for evolving  dark energy from these data.
The conclusions of the DESI team must therefore be a consequence of differences in the BAO and/or SN data.

\begin{figure}[!h]
\hspace{-0.1in}\includegraphics[width=2.7in]{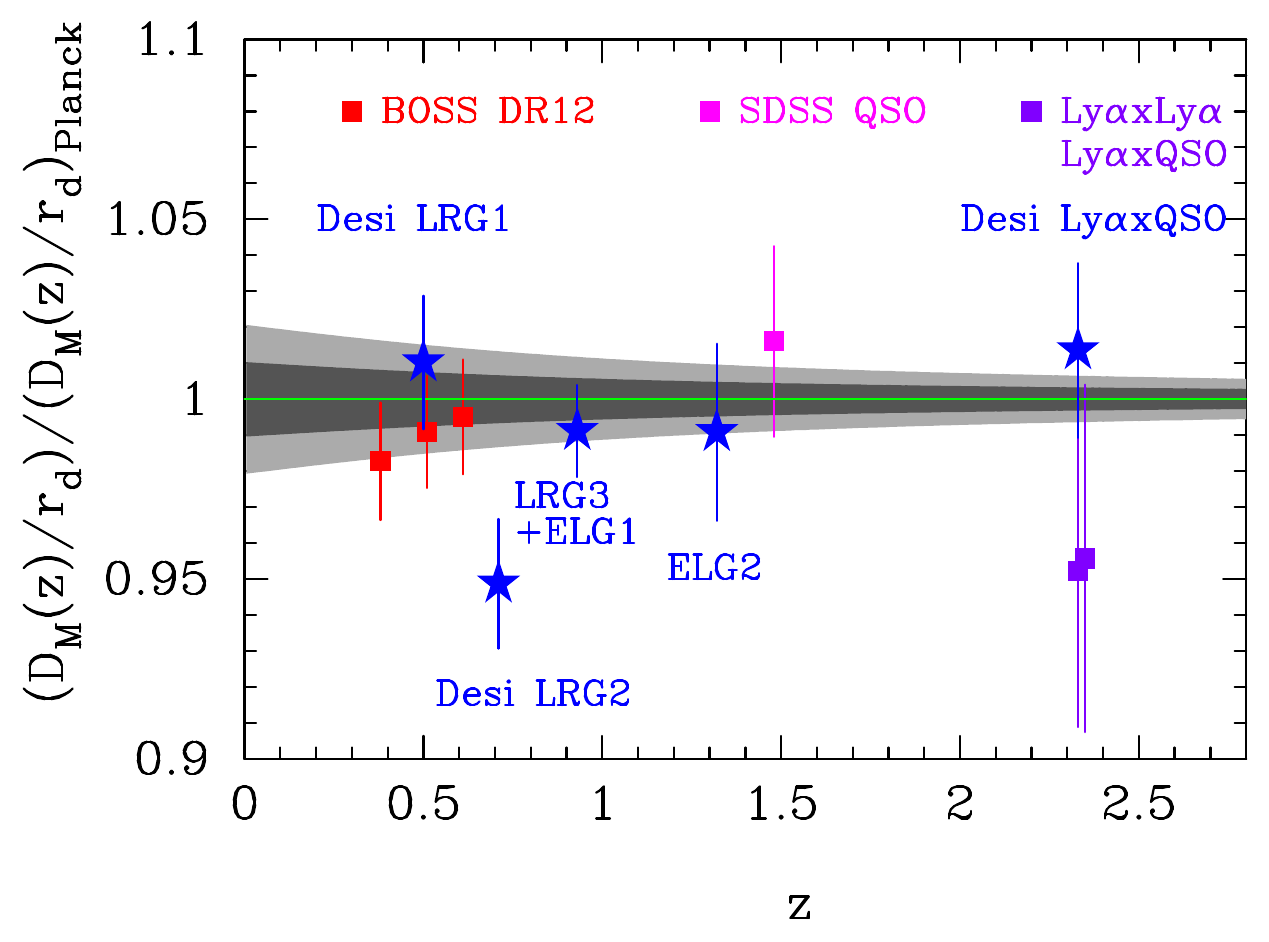}\hspace{0.0in}\includegraphics[width=2.7in]{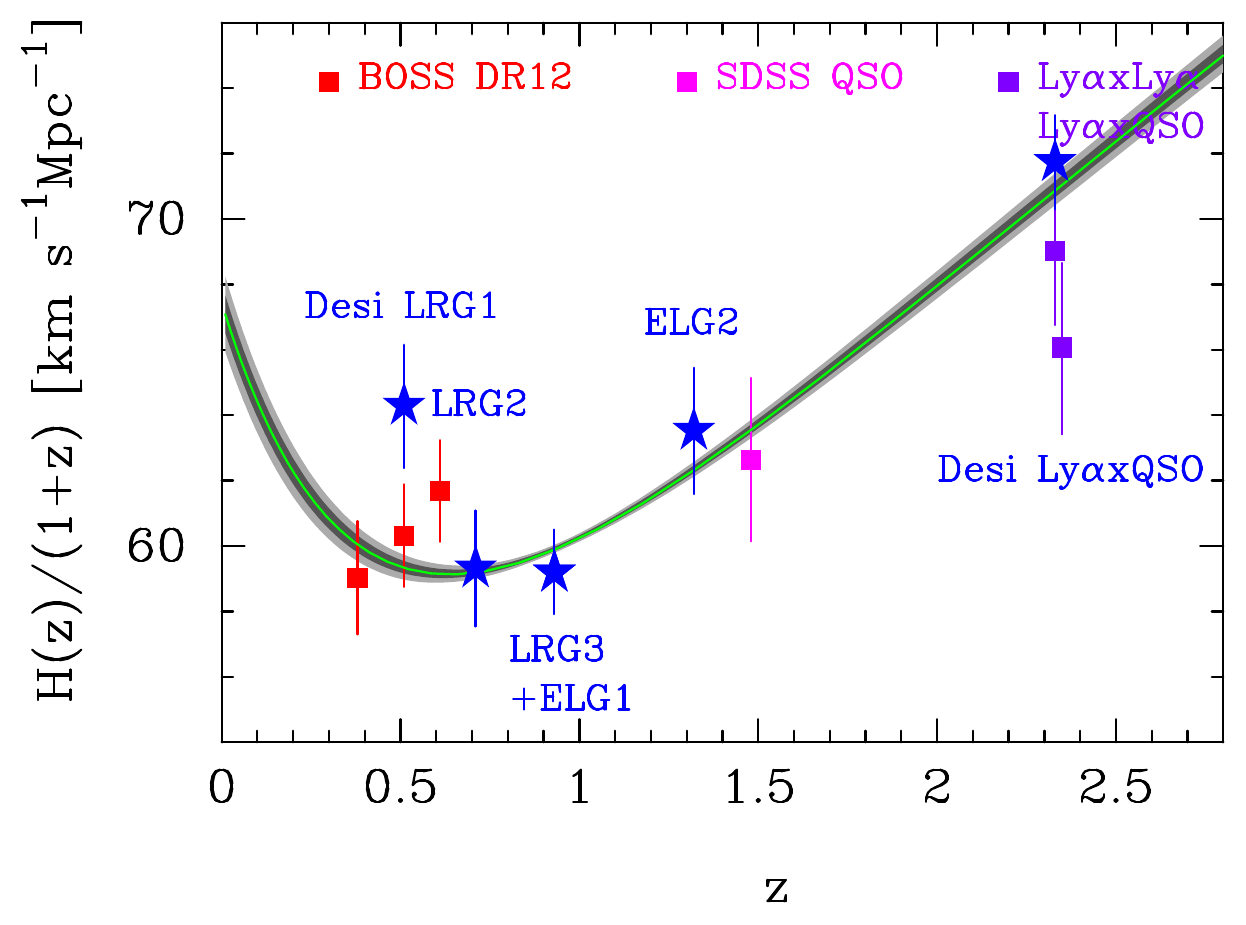}
\caption{Comparison of BOSS/SDSS and DESI measurements of $D_M(z)$ and $H(z)$. The green lines show the predictions of the
best fit \Planck\ base \LCDM\ cosmology and the grey bands show $1$ and $2\sigma$ errors. The BOSS/SDSS measurements
are from the following sources: BOSS DR12,  \cite{Alam:2017};  SDSS QSO,  \cite{Ata:2018}, BOSS Ly$\alpha\times$Ly$\alpha$, Ly$\alpha \times$QSO, \cite{deSainte:2019, Blomqvist:2019}. The DESI measurements are from \cite{DESI_1:2024, DESI_2:2024}.}
\label{fig:desi}
\end{figure}

Figure \ref{fig:desi} compares the DESI BAO results with the earlier
BOSS/SDSS measurements with
the predictions of the \Planck\ base \LCDM\ model. I plot only
measurements which have high enough signal to noise to measure both $H(z)$
and $D_M(z)$ from the BAO scale along and perpendicular to the
line-of-sight. The BOSS/SDSS measurements are in 
good agreement with the \Planck\ model. However, as noted by the DESI
team, there are two outliers amongst the DESI measurements; these are
$D_M$ for the DESI LRG2 sample ($\langle z \rangle = 0.71$), which
sits low compared to the \Planck\ prediction by $\sim 2.6\sigma$ and
$H(z)$ for the DESI LRG1 sample ($\langle z \rangle = 0.51$), which
sits high compared to the \Planck\ prediction by $\sim 2.8\sigma$. As a rough guide, if the errors are assumed to be Gaussian
and uncorrelated, the probability of getting two such deviant points out of $22$ points is about 1.8\%. This is unusual, but not excessively so. Importantly,  the two deviant DESI points are in tension  with  the results from BOSS RD12 
which measure $D_M$ and $H_z$ more accurately than DESI at similar redshifts\footnote{In contrast, the DESI Ly$\alpha \times$QSO measurement improves significantly on the 
earlier BOSS/SDSS constraint, and comes into even better agreement with the \Planck\ \LCDM\ model.}. Furthermore, the new
data points do not reinforce any coherent pattern in the earler BAO measurements that might indicate  a deviation from \LCDM, suggesting that the DESI outliers are just statistical fluctuations.  

The results presented in \citep{DESI_3:2024} strongly suggest that the
evidence for evolving dark energy is driven by the new SN
catalogues. As with the CMB, the DESI BAO measurements are strongly
degenerate in the parameters $w_0-w_a$ and show no significant
preference for evolving dark energy\footnote{The DESI team use $\Delta
  \chi^2_{\rm MAP}$ as a tension metric, which is the difference in
  $\chi^2$ for the maximum posterior allowing $w_0$ and $w_a$ to vary
  relative to $\chi^2$ for \LCDM\ (i.e. $w_0 = -1$, $w_a = 0$. For DESI
  BAO, they find $\Delta \chi^2_{\rm MAP} = -3.7$.}. It is only when the supernova samples
are added to \Planck\ CMB and  DESI BAO that they see a pull towards evolving dark energy, finding
a preference for evolving dark energy compared to \LCDM\ (using $\Delta \chi^2_{\rm MAP}$ as a tension metric)
 at about the $2.5\sigma$  (Pantheon+), $3.5 \sigma$ (Union 3, \cite{Rubin:2023}) and $3.8\sigma$ level (DESY5 SN, 
 \cite{Abbott:2024}).
 
 In summary, it is important to scrutinize the SN samples (particularly the new Union 3 and DESY5 samples)   to rule out the possibility that the DESI `dark energy tension' is caused by systematic errors in the SN data.\footnote{After the first
 version of this paper was submitted I compared the photometry of  SN  common to both the  Pantheon+ and the DESY5 catalogues.
 The results suggest that the pull towards evolving dark energy using DES5Y may be a consequence of a systematic error in
 matching the photometry of nearby SN with the more distant SN observed by DES \citep{Efstathiou:2024}.}

\section{Conclusions}
\label{sec:conclusions}

The six parameter \LCDM\ cosmology is remarkably successful. Yet we have little understanding at a fundamental level of the three key features of the model -- inflation, dark matter and dark energy. It is therefore possible that the tensions discussed in this article are caused  by  new physics. In this article, I have emphasised the fact that the six parameter \LCDM\ cosmology 
agrees to high precision with observations of the CMB anisotropies and with CMB lensing.
Observational evidence for departures from \LCDM\ is therefore conditional on the fidelity of
other types of astrophysical data. Fortunately, there are many new projects underway
that should not only clarify the tensions discussed here but should provide stringent 
new tests of the \LCDM\ cosmology.

\ack{I am indebted to my collaborators, Alex Amon, Steven Gratton, Calvin Preston, Vivian Poulin and Erik Rosenberg  for their contributions to some of the work described here. I thank Suhail Dhawan and Hiranya Peiris for discussions on Section \ref{sec:desi}. I thank the Leverhulme Foundation for the award of a Leverhulme Emeritus Fellowship.}


\bibliographystyle{mnras}
\bibliography{Challenges} 

\end{document}